# Towards a unified mechanistic understanding of the electrical response of bipolar membrane systems

Ayelet Ben-Kish Sharvit and Yoav Green*

Department of Mechanical Engineering, Ben-Gurion University of the Negev, Beer-Sheva 8410501, Israel

* Email: yoavgreen@bgu.ac.il

ORCID numbers: A.B-K.S. : 0009-0004-6365-9395

Y.G. : 0000-0002-0809-6575

**Abstract.**

Bipolar nanoporous membranes/nanochannels are used in water desalination and energy-harvesting systems that provide clean water and green energy, respectively. The ever-growing need requires that the performance of these systems be constantly improved. Since the underlying physics are still not fully understood, the empirical optimization process remains slow and inefficient. Utilizing a combination of theoretical tools and numerical simulations, we created a novel and robust framework to improve the design of these systems. We show that the system's response is determined by an interplay of the applied voltage and a parameter ($\eta$) that depends on the geometry and surface charge densities of both charged regions. At low voltages, we show that the response is mostly determined by $\eta$ such that the $\eta$-dependence can be represented by a simplified phase space. At high voltages, we show that this phase space becomes oversimplified. To demonstrate this, we conduct a systematic scan of this phase space, which spans a range of membrane configurations, from unipolar channels featuring a single charged region to bipolar channels composed of positively and negatively charged segments. We compare our numerically simulated current-voltage responses to those predicted by three known existing theoretical models, which are limiting scenarios within the phase space, and explain the observed deviations. Our findings underscore the need for a more comprehensive framework to capture the broad range of observed behaviors. Nonetheless, these same findings can be utilized to reduce the time/resource-intensive optimization process and improve the data interpretation of experiments and simulations of these systems.

## Graphical abstract

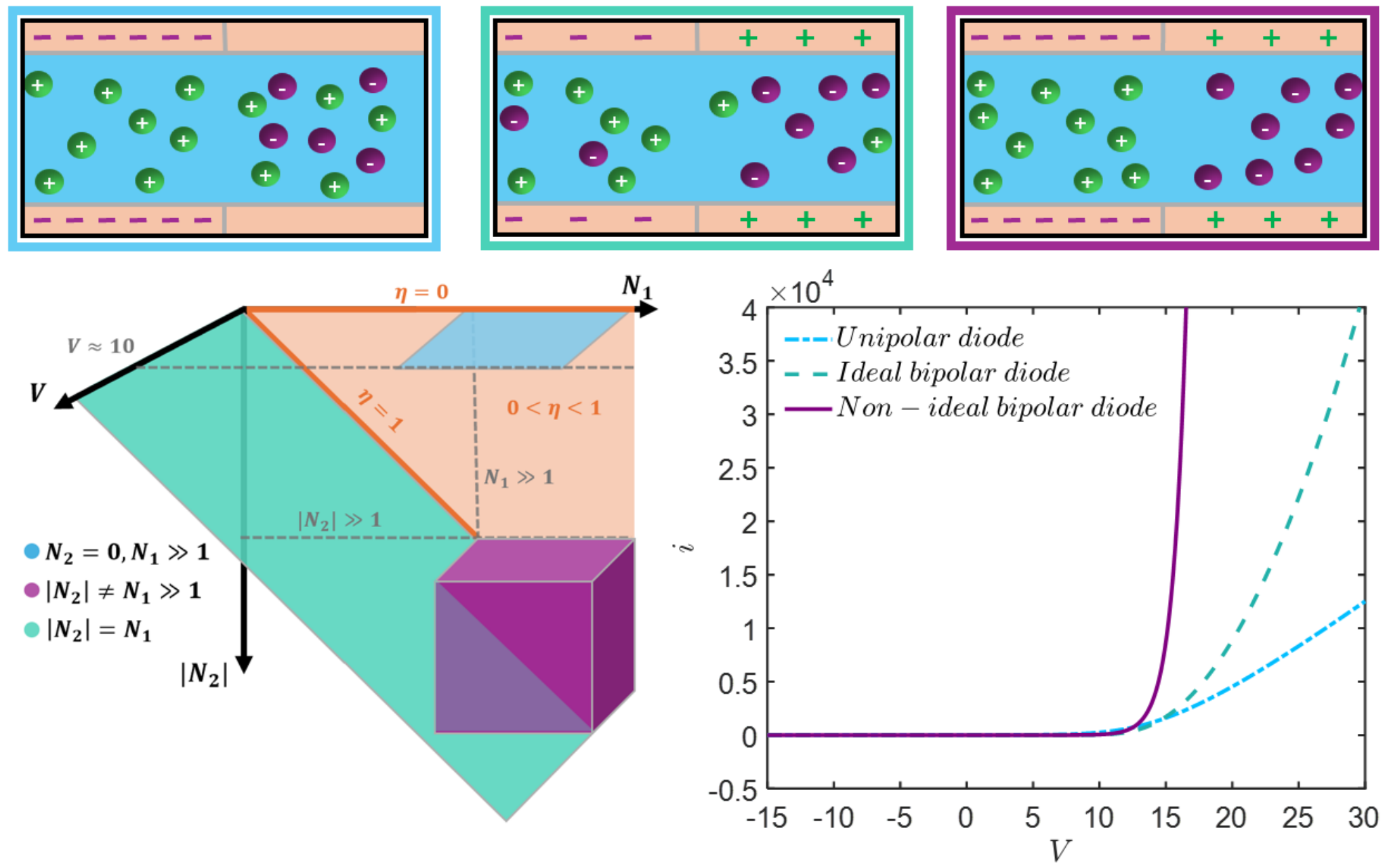

## 1. Introduction

Nanofluidic systems, particularly those based on ion-selective nanoporous membranes[1–10], have emerged as key platforms for controlling ionic transport at the nanoscale. Their ability to modulate ion transport[11–13] with high precision makes them central to various applications, such as water desalination[14–17], energy harvesting technologies[18–24], biosensing[25–28], and fluid-based electric circuits[29–31]. Additionally, these systems can be used for inorganic acid and base production from salt solutions[32], pH regulation, and recovery of resources from waste streams[33,34], with a capability for lithium and boron recovery[35,36], $CO_2$ capture[37,38], and recovery of ammonia from wastewater[39].

A central mechanism enabling such control is permselectivity, a symmetry-breaking property that allows preferential transport of ions with a charge of one sign over those of the opposite sign. This selective transport typically arises from engineered asymmetries in surface charge distributions or channel geometries[40–43]. This selective behavior, crucial for processes like ion separation or osmotic energy conversion, is often realized by engineering the distribution of surface charges along the channel walls. In particular, spatial variations in surface charge density can give rise to diode-like behavior[10,20,31,44–52], wherein the nanofluidic system rectifies ionic current analogous to solid-state diodes.

When an external electric field is applied, permselective systems exhibit directional ion transport, leading to spatial concentration gradients. These gradients are governed by the system's steady-state current-voltage ($I-V$) response, which reflects the coupled influence of geometry, bulk ion concentration, and surface charge characteristics. The development of such gradients, along with their electrical signature, is commonly called concentration polarization[41–43,53–55], and can be found in virtually all nanofluidic systems. Its known characteristics provide the background needed to characterize and analyze the system properties and response.

Among the simplest and most studied configurations are bipolar (Figure 1) nanochannels with either uniform or conical geometries, where rectification efficiency can reach several orders of magnitude depending on the interplay between charge distribution and geometry.

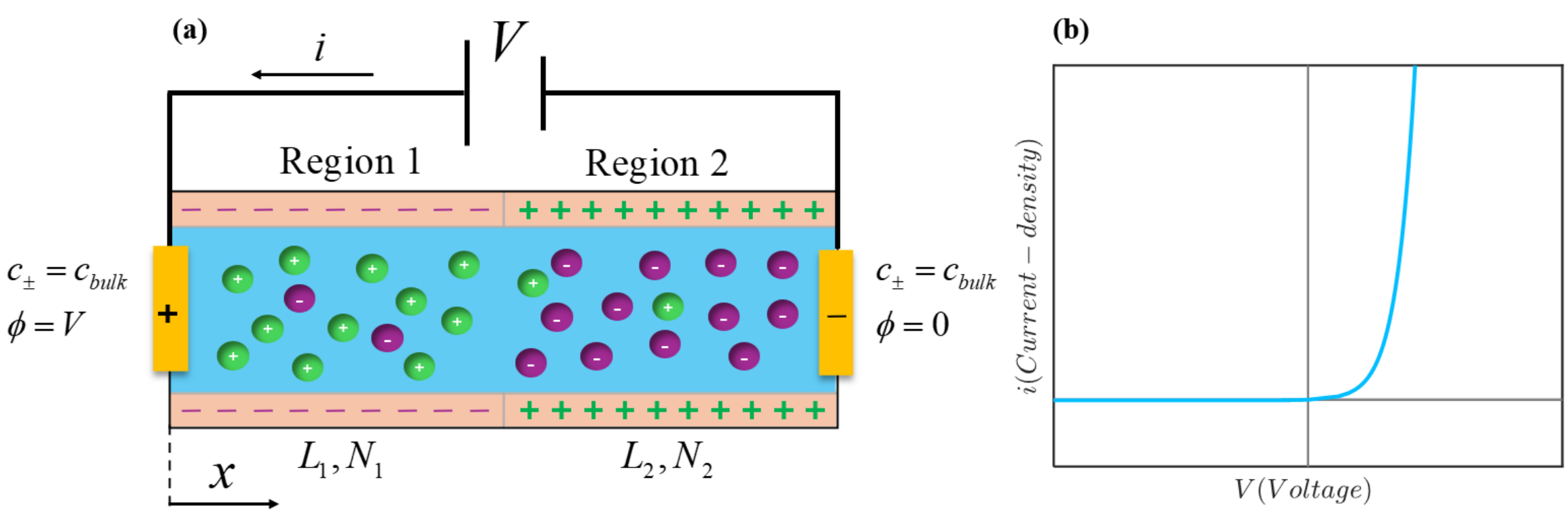

**Figure 1**. (a) Schematic of the 1D system, comprised of two differently charged regions, referred to as "Region 1" and "Region 2", and defined by $L_1, N_1$ and $L_2, N_2$ respectively. (b) Schematic bipolar current-density-voltage curve.

Key studies, including Vlassiouk *et al*. (2008)[50], Green *et al*. (2018)[51], and Picallo *et al*. (2013)[52], derived three different $i-V$'s, have significantly advanced the theoretical understanding of bipolar systems, particularly revealing the asymmetric current-voltage ($i-V$) curve under specific conditions. Despite these contributions, a universal analytic solution for systems with varied surface charges outside large-scale limits remains elusive. In this work, we numerically investigate the electrical response of a wide range of bipolar systems while making a detailed comparison with the theoretical models. We demonstrate and explain why each model eventually fails. We also rationalize the behavior of the electric response of bipolar systems whose parameters do not align with those used in the three previous models. Thus, this work seeks to address a general knowledge gap and explore these uncharted regions in the large parameter space that can be used to design bipolar nanofluidic systems. .

This work is divided as follows. Section 2 presents the 1D model of the bipolar nanofluidic system. In particular, we shall define geometry, governing equations, and boundary conditions. In Sec. 3, we review the three known models[50–52] for the $i-V$'s. The underlying assumptions of each model are discussed, where particular emphasis is put on when each of these assumptions is expected to fail. In Sec. 4, we present the results of extensive numerical simulations, which allow us to verify while providing a mechanistic explanation when the underlying assumptions fail. Finally, we conclude our discussion in Sec. 5, where we provide an alternative approach to carefully use existing knowledge and discuss future work that needs to be undertaken to close the existing knowledge gap.

## 2. Model

Before proceeding, an essential comment regarding the notations is needed. In this work, we utilize the more general and robust non-dimensional formulation, which uses a reduced parameter set than a dimensional formulation. The former allows for comparison across different models/experiments characterized by the same non-dimensional parameters, even when the dimensional parameters differ. While we use the non-dimensional formulation, we shall provide the proper dimensionalizations throughout this work, and on certain occasions, we will also utilize the more intuitive dimensional parameters. By default, all non-dimensional variables and parameters are denoted without tildes, while all dimensional variables and parameters are denoted with tildes. For example, the non-dimensional and dimensional cross-sectional area is given by $S_{\text{nano}}$ and $\tilde{S}_{\text{nano}}$, respectively.

### 2.1 Geometry

We consider a 1D nanofluidic channel comprised of two regions with oppositely charged surface densities (Figure 1), containing a binary electrolyte (the properties of the electrolyte are discussed in the next subsection). The left region, denoted as region 1, has a negative surface charge, $\sigma_{s1}$, and length $L_1$. The right region, denoted as region 2, has a positive surface charge, $\sigma_{s2}$, and length $L_2$. The system is bounded at both ends by bulk reservoirs held at uniform/bulk concentration. An external potential drop $V$, positive when applied from left to right (the origin of the coordinate system is at the leftmost point of region 1), results in an electrical current, $I$, or electrical current density, $i = I / S_{\text{nano}}$ where $S_{\text{nano}}$ is the cross-sectional area. In the remainder of this work, we will refer to the current and current density interchangeably.

In 1D models, various transport characteristics in the longitudinal direction are taken by dividing a 3D quantity by the cross-sectional area, as was done above with $I$ and $i$. To reduce the dimension of the system from 3D or 2D to 1D, a similar process is taken for surface quantities such as the surface charge density. To account for the surface charge density, in a reduced system, the surface charge density is integrated along the perimeter of the channel, $\tilde{P}_{\text{nano}}$, and then divided $\tilde{S}_{\text{nano}}$ and the Faraday constant $\tilde{F}$

$$\tilde{N}_{k=1,2} = -\frac{\tilde{\sigma}_{\text{sk}}\tilde{P}_{\text{nano}}}{\tilde{F}\tilde{S}_{\text{nano}}}. \tag{1}$$

The resultant parameter, which has units of concentration, is often termed the average excess counterion concentration or fixed charge concentration. This concentration produces a surplus of ions of the opposite charge to the surface charge density. Naturally, the (dimensional) average excess concentration in regions 1 and 2 is denoted as $\tilde{N}_1$ and $\tilde{N}_2$, respectively.

It is important to emphasize that, in contrast to nanochannel systems, where $\tilde{N}_1$ and $\tilde{N}_2$ are given by the simple equation above (relating the excess counterion concentration to the surface charge), in membrane systems, a simple equation is not known. Like the nanochannel, the excess counterion concentration depends on the surface charge density, while the dependency on the spatially heterogeneous geometry is not known. Thus, the surface charge density cannot be extracted directly. Regardless of whether one can transition between $\tilde{N}_k$ and $\tilde{\sigma}_{sk}$, in both the simple nanochannel system and the more complicated nanoporous membrane system, the values for $\tilde{N}_1$ and $\tilde{N}_2$ are extracted from experiments.

### 2.2 Governing equations and boundary conditions

The non-dimensional equations governing 1D, steady-state, and convection-less ion transport through a permselective medium for a symmetric and binary electrolyte (such as KCl), where the diffusion coefficients of the positive and negative ions are the same, and the valences are $\pm 1$, are the Poisson-Nernst-Planck (PNP) equations

$$2\varepsilon^2 \phi_{,xx} = -\rho_e \tag{2}$$

$$-j_\pm = c_{\pm,x} \pm c_\pm \phi_{,x}, \tag{3}$$

where Eq. (2) is the Poisson equation for the non-dimensional electric potential $\phi$, which depends on the space charge density $\rho_e$ and the non-dimensional Debye layer, $\varepsilon$, described below. Eq. (3) is the Nernst-Planck equation for the positive and negative ionic fluxes, $j_\pm$, which, through ionic flux conservation, determines the distribution of the ionic concentrations $c_\pm$.

The space charge density in each region is defined as the difference between the positive and negative ion concentrations, while the contribution of the surface charge is effectively modeled by the average excess counterion concentration (in each region), $N_1 > 0$ and $N_2 < 0$ such that

$$\rho_e = \begin{cases} c_+ - c_- - N_1, 0 \le x \le L_1 \\ c_+ - c_- - N_2, L_1 \le x \le L_2 \end{cases}. \tag{4}$$

Here, it has been assumed that valences are of equal magnitude but opposite sign ($z_\pm = \pm 1$).

The normalizations used for Eqs. (2)–(4) are the following. The potential is normalized by the thermal potential, $\tilde{\phi}_{th} = \tilde{\Re}\tilde{T}/\tilde{F}$, where $\tilde{\Re}$ is the gas constant, $\tilde{T}$ is the absolute temperature, and (once more) $\tilde{F}$ is Faraday's constant. The spatial coordinates and derivatives are normalized by a characteristic length $\tilde{L}$, which can be either $\tilde{L}_1$ or $\tilde{L}_2$. The ionic concentrations $c_{\pm}$ as well as the average excess counterion concentrations $N_1$ and $N_2$ are normalized with the bulk concentration, $\tilde{c}_{\text{bulk}}$ such that $c_{\pm}=\tilde{c}_{\pm}/\tilde{c}_{bulk}$, $N_{k=1,2}=\tilde{N}_k/\tilde{c}_{bulk}$. Then the ionic fluxes $j_{\pm}$ are normalized by $\tilde{D}\tilde{c}_{bulk}/\tilde{L}$ where $\tilde{D}$ is the diffusion coefficient of both species. In Eq.(2), the normalized Debye length is introduced,

$$\varepsilon=\frac{\tilde{\lambda}_D}{\tilde{L}}=\frac{1}{\tilde{L}}\sqrt{\frac{\tilde{\varepsilon}_0\varepsilon_r\tilde{\Re}\tilde{T}}{2\tilde{F}^2\tilde{c}_{bulk}}} \tag{5}$$

where, $\tilde{\lambda}_D$ is the dimensional Debye length, which depends on the permittivity of vacuum, $\tilde{\varepsilon}_0$ , and the relative permittivity of the electrolyte, $\varepsilon_r$ .

Finally, after providing the normalizations, we can mathematically define the boundary conditions that were briefly discussed above. The boundary conditions for a system held at symmetric concentrations at the two ends of the system and subjected to a potential drop $V$ are given by

$$c_{\pm}(x=0)=1 \quad , \quad c_{\pm}(x=L_1+L_2)=1 \quad , \quad \phi(x=0)=V \quad , \quad \phi(x=L_1+L_2)=0\,. \tag{6}$$

The boundary condition for the potential is rather intuitive – the drop across the system yields an electric field across the system that drives the ions, and based on the application, can also electrophoretically drive particles, colloids, and proteins. The boundary of a symmetric concentration is more nuanced. There are some applications, such as biosensing and fluid-based electric circuits, where the desired result is detection and/or current rectification. There, the use of symmetric concentrations simplifies the analysis (by removing an additional parameter that can affect the overall response). There are other applications, like water desalination, energy-harvesting, and mineral extraction, where the systems are subjected to asymmetric salt concentrations such that our selected boundary condition is not the most appropriate. Nonetheless, there is good reason for selecting this boundary condition – this is the ability to compare simulations and all the theoretical models. Most of the theoretical models presented below were derived with this same boundary condition. The exception is the work of Picallo *et al.*[52] who derived their $i-V$ for asymmetric salt concentrations. However, like all the models, due to its inherent assumption, this model is also limited to a small parameter space. For us to show the limitations of all models under the same conditions, and under the assumption that if the model fails for symmetric concentration, it will also fail for asymmetric concentration, we have restricted our discussion to the case of symmetric concentration. In recent years, we have published several works[13,56–59] on the electrical response of a single charged region ("regular/standard" membrane/nanochannel) subjected to asymmetric salt concentrations. We believe that eventually the formulation of all these works can be merged to model the electrical response of bipolar systems subjected to asymmetric salt concentrations. However, this is left for future work.

Once Eqs. (2)–(4), subject to Eq. (6), are solved for the potential and concentrations, one can also calculate the fluxes $j_+$ and $j_-$. From these fluxes, one can calculate the electrical current density, $i$, and the salt current density, $j$, given by

$$i = j_+ - j_- \quad , \quad j = j_+ + j_- . \tag{7}$$

Note that $\tilde{i}$ and $\tilde{j}$ have different dimensional units: $\tilde{j}$ has units of $\tilde{D}\tilde{c}_{bulk} / \tilde{L}$ while $\tilde{i}$ has units of $\tilde{F}\tilde{D}\tilde{c}_{bulk} / \tilde{L}$.

One last comment is needed regarding the effects of the Debye length $\tilde{\lambda}_D$ and its various normalizations. For a nanochannel, as the one schematically shown in Figure 1(a), in principle, $\tilde{\lambda}_D$ can be normalized by either the length or the height/radius. For simplicity, we can denote these two normalized quantities by $\varepsilon_L$ and $\varepsilon_a$, respectively. In our work, $\varepsilon = \varepsilon_L$ appears explicitly [Eqs. (2) and (5)], while $\varepsilon_a$ is implicit. This will be demonstrated for the case of a circular nanochannel with $\tilde{P}_{\text{nano}} = 2\pi\tilde{a}$ and $\tilde{S}_{\text{nano}} = \pi\tilde{a}^2$. Then, the non-dimensional excess counterion concentrations can be written as $N_{1,2} = -4\sigma_{sk}\varepsilon_a^2,\ k = 1,2$ (where $\sigma_{sk}$ has been normalized by $\tilde{\varepsilon}_0\varepsilon_r\tilde{\phi}_{th} / \tilde{a}$ [56,60]). Thus, the values of $N_{1,2}$ implicitly account for any Debye length overlap (small or large) in the lateral direction, along with the effects of the surface charge.

The two normalized Debye lengths are not independent, but are related by the ratio $\varepsilon_L / \varepsilon_a = a / L$. In this work, we assume that the length is substantially larger than the radius, such that the assumption $\varepsilon = \varepsilon_L \ll 1$ always holds. One can always relax such an assumption (theoretically, numerically, or experimentally). For example, in experiments, one can find $\varepsilon_L = 1$ for an ultra-thin membrane. In this work, we do not consider such a scenario for several reasons. First, on the practical level, all the theoretical $i - V$'s presented below implicitly assume that $\varepsilon_L \ll 1$ by arguing that local electroneutrality can be imposed. If this assumption no longer holds, then electroneutrality is questionable. Thus, to conduct a detailed comparison with theory, we need our numerical parameter space to match that of the theory. Second (which is also partially related to the first argument), we will utilize the sharp Donnan potential drops to rationalize the failure of the models. If $\varepsilon_L \sim 1$, the Donnan potential drops no longer need to be sharp, but will occur over the entire channel, negating our ability to rationalize theory. Thus, since theory imposes in the limit of $\varepsilon \to 0$, so do we. Nonetheless, this work can be considered the first step in understanding the response of bipolar systems as a function of $\varepsilon = \varepsilon_L$. Here we have addressed $\varepsilon_L \ll 1$. Future works can address $\varepsilon_L \sim 1$ and $\varepsilon_L \gg 1$.

### 3. Current-voltage response

The resultant current-density-voltage, $i-V$, response strongly depends on all the remaining system parameters ($L_1$, $L_2$, $N_1$, and $N_2$). Past works have shown that the dependency can be reduced to depend on a single (non-dimensional) control parameter that accounts for all four parameters at once[40,44,45,61]

$$\eta = \frac{L_2}{L_1}\frac{|N_2|}{N_1}. \tag{8}$$

In the following, we will demonstrate how this parameter can be divided into three subsets, and how the $i-V$ varies with this subset. However, before that, three last simplifying comments are needed. First, Eq. (8) depends on four parameters, and as such, in principle, a four-dimensional parameter space is needed to fully characterize such a system. In this work, we shall mostly consider and present results for the case $L_1 = L_2$, and reduce the phase space to 2D. A schematic of this phase space is shown in Figure 2(a). Even though we focus on the simplified $L_1 = L_2$ scenario, one can remove this assumption and find that the qualitative nature of the results presented in this work remains unchanged. Second, since the governing equations for ion transport across a bipolar nanochannel or a bipolar membrane are the same, once $N_1$ and $N_2$ have been determined, via experiments the analysis utilizing Eq. (8) holds equally for either system. Third, one can interpret $\eta$ to be the ratio of the total excess counterion charge in each ration. In a 2D or 3D system[40], the ratio is not of the total but rather of the average counterion charge.

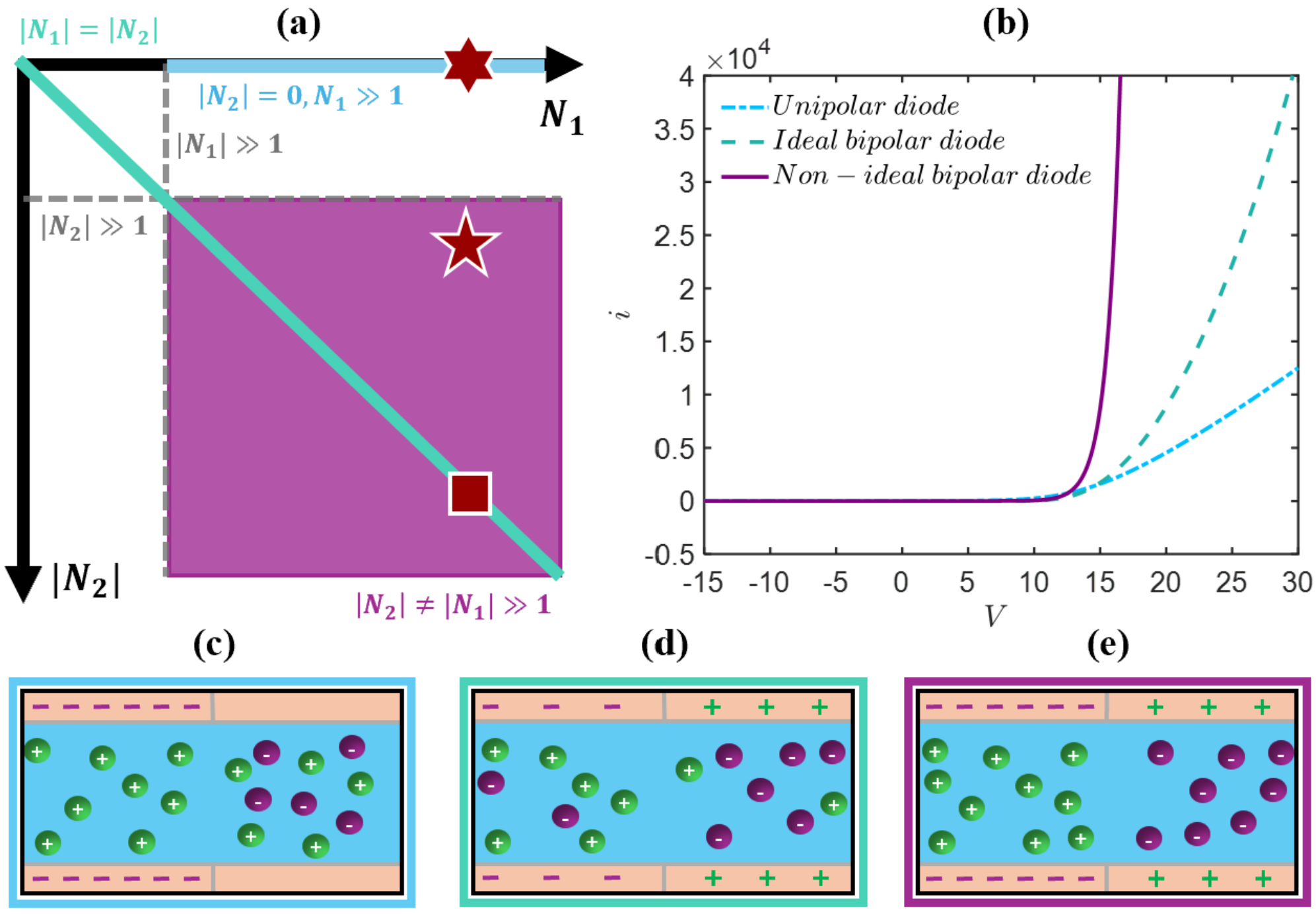

**Figure 2.** (a) $N_1 - N_2$ phase space diagram denoting the three known current-density-voltage, $i-V$, responses for bipolar nanofluidic systems. (b) The $i-V$ curves for the three known models. (c)–(e) depict the three respective systems, and their color coding that will be used throughout this work to denote these scenarios: (c) unipolar diode, (d) ideal bipolar diode, and (e) non-ideal bipolar diode.

This dimensionless parameter, $\eta$, can be divided into three scenarios. In what follows, we provide the subdivision and the rationality behind the selected name/terminology. After the presentation of the division, we provide a detailed discussion regarding the respective $i-V$ of each scenario [Figure 2(b)]. The scenarios are:

- $\eta = 0$: The unipolar configuration, in which there is only one charged region, whereby $N_2 = 0$ [Figure 2(c) and the dot-dashed blue line in Figure 2(b)].
- $\eta = 1$: The "ideal bipolar" configuration, where the two regions have equal and opposite counterion concentration $N_1 = |N_2|$ [Figure 2(d) and the dashed turquoise line in Figure 2(b)].
- $0 < \eta < 1$: Non-ideal bipolar configuration, where both regions are charged, though not necessarily of the same magnitude [Figure 2(e) and the purple line in Figure 2(b)]. We already note here that the existing model for this scenario will assume that $N_1 \gg 1$ and $|N_2| \gg 1$.

It is important to note that in all three scenarios, the $i-V$ behavior is diode-like: the current increases unhindered on one side but is hindered/limited on the other side [Figure 2(b)]. In this manner, all three systems are "diodes". However, the name 'diode' also carries subtle differences. The unipolar diode[20,51,62] is the simplest to understand – there is only one charged region and hence it is not bipolar, yet it still has a diode-like response. In contrast, the other two scenarios are bipolar diodes. In a previous work[44], we termed the $\eta=1$ scenario as a "symmetric bipolar" system and the $\eta\neq 1$ scenario as an "asymmetric bipolar" system. However, if one considers their $i-V$'s – both are asymmetric, and thus we found the terminology misleading. Since then[44,61], we have replaced the names to be "ideal" and "non-ideal" bipolar diode systems. Here, there is nothing "ideal" or "non-ideal" about these systems. They are just the simplest set of names we found useful to differentiate between all three systems.

In the following, we review each of the known models and their limitations. We shall start with the simplest of the three models – the unipolar diode. Then we will discuss the ideal bipolar diode, and finally, the non-ideal bipolar diode. It is important to emphasize that in all scenarios, it is assumed that $\varepsilon \ll 1$ such that the system is electroneutral (such that $\rho_e=0$). In each scenario and each of the regions, this yields different results.

### 3.1 Unipolar diode

The $i-V$ for the case of the unipolar diode scenario, wherein $N_2=0$, shown as the dot-dashed blue colored line in Figure 2(b), is given by[51]

$$V=\frac{iL_1}{N_1}+2\ln\left(1+\frac{iL_2}{2}\right). \tag{9}$$

Equation (9) is comprised of two terms. The first term represents a linear/Ohmic voltage drop across the charged region [region 1 in Figure 2(c)]. The second term corresponds to the drop across region 2. Here, we can see that for positive voltages, the current increases unhindered. At large currents, the second term is still much smaller than the first term [$O(i)\gg O(\ln(i))$]. As a result, at high voltages, the current is approximately linear. At negative currents, the second dominates the response, and there is a limiting current ($i_{\lim}=-2/L_2$) which the current cannot surpass (in the absolute sense).

To understand this diode-like result and the limitations of this model, it is essential to discuss the embedded assumptions of this model, so that later we will be able to show when these assumptions fail. It is assumed that the left side, which is charged, is ideally selective.

Ideal selectivity is implemented by asserting two things (the order of which is not important). First, in region 1, there is no negative ion transport, $j_- \equiv 0$. However, 1D flux conservation then requires that $j_- \equiv 0$ in region 2. Thus, $j_- \equiv 0$ in the entire system. Then, based on Eq. (7), we find that $i = j = j_+$ everywhere in the system. Second, when implementing ideal selectivity, one also typically needs to require that $N_1 \gg 1$. This will lead to spatially independent concentration profiles wherein $c_{1+} \simeq N_1$ and $c_{1-} = N_1^{-1}$. Soon, it will be shown that both assumptions fail at sufficiently large voltages/currents.

### 3.2 Ideal bipolar diode

The $i-V$ for the case of the ideal bipolar diode scenario, wherein $\eta = 1$, shown as the dashed turquoise-colored line in Figure 2(b), is given by[34,40,50]

$$V = \left( \frac{P_2}{N_2} - \frac{Q_{-,2}}{N_2} - \frac{P_1}{N_1} + \frac{Q_{+,1}}{N_1} \right) + \ln \left[ \left( \frac{Q_{-,2} + N_2}{P_2 + N_2} \right) \left( \frac{P_1 + N_1}{Q_{+,1} + N_1} \right) \right] \tag{10}$$

where

$$Q_{\pm,k} = \sqrt{4 \pm 2iN_k L_k + N_k^{\;2}}, \quad P_k = \sqrt{4 + N_k^{\;2}}, \text{ where } k = 1,2. \tag{11}$$

Here, the $P_k$ functions are the terms that account for the Donnan potential, while the $Q_{\pm k}$ functions account for the internal concentration polarization in the two regions. Note that Vlassoiuk and Siwy[50] and Green *et al.*[40] provide slightly different expressions for Eq.(10). In the earlier work of Vlassoiuk and Siwy[50], it was assumed that $L = L_1 = L_2$ such that $N_1 = |N_2|$. Both simplifications result in a simpler final expression. These "restrictive" assumption was later removed by Green *et al.*[40] to yield the more general Eq.(10) provided here. In fact, Ref.[40] considered a 2D system where the heights of regions 1 and 2 were not the same, and thus provided a more general expression for $\eta$ [Eq.(8)], $Q_{\pm,k}$ and $P_k$ [Eq. (11)].

The mathematical analysis of Eq. (10) is not as straightforward as in Eq. (9), but the general characteristics can easily be understood. Similarly to Eq. (9), Eq. (10) also consists of two distinct terms. However, in contrast to Eq. (9) where the first term is associated with region 1 and the second term is associated with region 2, here both terms depend on the properties of both regions. The first term, which depends on the square root of the current, can grow unhindered for positive currents and voltages. For negative voltages, a limiting current can be calculated. However, this limiting current is less restrictive than the limiting current calculated from the second logarithmic term [$i_{\lim} = 2/(N_2 L_2)$, where $N_2 < 0$, for negative voltages][40].

To derive Eq. (10), it is assumed that $j = 0$. This leads to positive and negative fluxes being antisymmetric (such that $j_+ = -j_-$). Then, Eq. (7) yields $i = 2j_+$. Another signature of this model is that the resultant concentration and electric potential distributions always have a square-root dependency on $x$. Any deviation from $\eta = 1$, will lead to a breakdown of this theory.

### 3.3 Non-ideal bipolar diode

The third model presented corresponds to a configuration in which the two nanochannels possess surface charges of opposite sign but not equal in magnitude. In both regions is assumed that the excess counterion concentrations are large $(N_1, |N_2| \gg 1)$, such that each region is approximately "ideally" selective (a thorough discussion regarding the region's selectivity and the system's selectivity will be provided later). The $i-V$ for this scenario, termed the non-ideal bipolar diode, is shown as the purple colored line in Figure 2(b), and is given by[52]

$$V = i\left(\frac{L_1}{N_1} - \frac{L_2}{N_2}\right) + \ln\left[1 - \frac{i}{2}L_2N_2\left(1 + \frac{L_2N_2 + L_1N_1}{L_1N_1 - L_2N_2}\right)\right]. \tag{12}$$

Several comments regarding Eq.(12) are needed. In Ref.[52], the roles of regions 1 and 2 are reversed (i.e., their region 1 is positive and region 2 is negative), and their potential is defined as positive from right to left. Also, Ref.[52] assumes that the lengths of both regions are the same (specifically, they assume $L_1 = L_2 = 1/2$). In addition, in their system, they have $N_1 > 0$ and $N_2 > 0$, where an additional minus is added before one of them (here, for $N_2$). Equation (12) represents the $i-V$ system as shown in Figure 1(a) for $N_1 > 0$ and $N_2 < 0$, and arbitrary lengths.

Equation (12) has a similar form to the previous models. The first term is a linear Ohmic term that has unhindered growth for positive voltages. The logarithmic term will yield a negative limiting current for negative voltages (given in Table 1).

In the derivation of this model, it is assumed that $N_1 \gg 1$ and $|N_2| \gg 1$, such that both region 1 and region 2 are "almost" ideally selective. However, the implementation of the selectivity differs from that used in the unipolar model. In region 1 of the unipolar scenario, it was assumed that: $j_- \equiv 0$, $c_{1+} \simeq N_1$ and $c_{1-} = N_1^{-1}$. Here, it is assumed that $j_-$ is a small but non-zero ($j_- \neq 0$) such that $c_{1+} \simeq N_1$ and $c_{1-} \simeq N_1^{-1} - j_- x$. In region 2, we have a similar behavior where $c_{2-} \simeq N_2$ and $c_{2+} \simeq N_2^{-1} + j_+(L_1 + L_2 - x)$. Importantly, $j_+ \neq j_-$ such that $j_- \neq 0$. As such, the response is not antisymmetric. Ref.[52] notes that the $i-V$ fails at high voltages. However, an explanation for this failure is not provided. We will provide this explanation shortly.

One last comment is needed. Equation (12) is written here as it was derived and originally presented in Ref.[52]. However, the authors of Ref.[52] argue that under certain conditions, one can neglect the Ohmic term. To ensure our analysis is complete, we analyze the results of Eq. (12) twice, once with the linear Ohmic term and once without it. Both models are essentially the same, except that the model without the linear term makes one additional assumption, thereby decreasing its robustness. Both curves will be plotted with purple lines. The curve without the Ohmic term will be denoted with a solid line, while the curves with the Ohmic term will be denoted with a dotted line. As we will shortly show, both models will eventually fail at high voltages. Nonetheless, it is important to note that for negative voltages, when limiting currents appear, the Ohmic term does not affect the limiting currents. In other words, the limiting current depends solely on the logarithmic term.

### 3.4 $i-V$ summary

Thus far, we have discussed the three known models for $i-V$ bipolar systems. These can be summarized in Table 1. In particular, we provide here, for the first time, a comprehensive comparison of the low-voltage-low-current Ohmic conductance density ($g = i/V$ when $i \ll 1$) and limiting current of all three systems. Note that the Ohmic conductance is related to the Ohmic conductance density $G = gS_{nano}$. As we will discuss later, in the Discussion section (Sec. 5), all these expressions are immensely important for experimentalists to ascertain what is the most appropriate model to use when comparing to theoretical models.

**Table 1**. Ohmic conductance and limiting current for each of the three models analyzed.

| Model Name | Equation number | Ohmic conductance | Limiting current |
|---|---|---|---|
| Unipolar diode | Eq. (9) | $g_{ohm} = \frac{L_1}{N_1} + L_2$ | $i_{\lim} = -\frac{2}{L_2}$ |
| Ideal bipolar diode | Eq. (10) | $g_{ohm} = \frac{L_1}{4}(-N_1 + \sqrt{4+N_1^2}) + \frac{L_2}{4}(-N_2 + \sqrt{4+N_2^2})$ | $i_{\lim} = \frac{2}{L_2 N_2}$ |
| Non-ideal bipolar diode | Eq.(12) | $g_{ohm} = \frac{L_1}{N_1} - \frac{L_2}{N_2} + \left( \frac{1}{N_1 L_1} - \frac{1}{N_2 L_2} \right)^{-1}$ | $i_{\lim} = \frac{1}{N_2 L_2} - \frac{1}{N_1 L_1}$ |

## 4. Results

As we have previously remarked, in general, the $i-V$ depends on $L_1$, $L_2$, $N_1$, and $N_2$. For the sake of simplicity, in our analysis, we will focus on the case that $L_{1,2}=1$, while varying $N_1$, and $N_2$. Thus, it reduces the phase space to essentially 2D. Figure 3 presents this reduced $N_1-N_2$ phase space. The six markers denoted in this figure correspond to the system configurations considered in this work, and whose values are detailed in Table 2. Most of the considered scenarios utilize a fixed $N_1=1000$, while $N_2$ is varied (see the black dashed line). However, there is also one scenario in which $N_1=|N_2|=100$ (triangle marker). Finally, Table 2 summarizes the parameters used for the numerical simulations. The details of the numerical simulations can be found in the Associated Content (Methods).

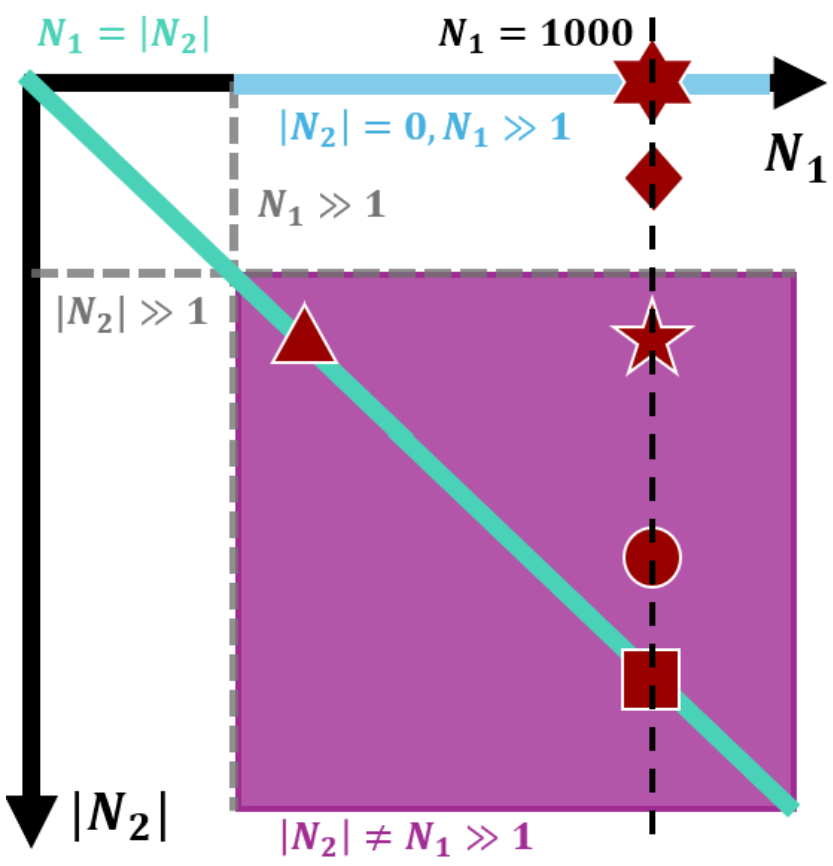

**Figure 3.** Simplified $N_1-N_2$ phase space diagram with six markers corresponding to the various scenarios detailed in Table 2.

**Table 2.** Values for the excess counterion concentrations $N_1$ and $N_2$ used in simulations shown in Figure 4. The markers correspond to those shown in Figure 3.

| Figure No. | $N_1$ | $N_2$ | Marker |
|---|---|---|---|
| Figure 4(a) | 100 | -100 | ▲ |
| Figure 4(b) | 1000 | -1000 | ■ |
| Figure 4(c) | 1000 | -600 | ● |
| Figure 4(d) | 1000 | -100 | ★ |
| Figure 4(e) | 1000 | -10 | ◆ |
| Figure 4(f) | 1000 | 0 | ✡ |

### 4.1 Current-voltage analysis

Figure 4 presents the $i-V$'s for the scenarios denoted in Figure 3 (whose details are provided in Table 2). In what follows, we will analyze and discuss each of the scenarios and compare them to the relevant theoretical model(s).

First, in Figure 4(a–b), we consider the two scenarios of $N_1=-N_2=100$ and $N_1=-N_2=1000$, which corresponds to the case of $\eta=1$. Here, one should expect the ideal bipolar model to have the best correspondence. Indeed, this is what we find. However, since in both scenarios $N_1 \gg 1$ and $|N_2| \gg 1$, we would also expect the non-ideal bipolar scenario to have equally good correspondence, though, as can be observed, this is not the case. The non-ideal bipolar model has relatively good correspondence at low voltages but fails at high voltages. This failure, which depends on both values of $N_1$ and $N_2$ will be discussed thoroughly in the sub-sections.

From this point forward, we shall consider only $N_1=1000$, while we vary the value of $N_2$. In Figure 4(c), we consider the case of $|N_2|=600 \gg 1$. Here, one would expect the non-ideal bipolar model to have the best fit. As can be observed in the inset, there is very good correspondence at low voltages. However, here too, at high voltages, the models fail. Surprisingly, the ideal bipolar model has relatively good (but not excellent) correspondence for all voltages. The reason this occurs is that even though the model is not "symmetric", the salt current density is approximately zero, $j \simeq 0$, which is one of the underlying assumptions of Eq. (10). That $j$ is not zero, but "relatively" close to zero, will be demonstrated shortly (Sec. 4.4).

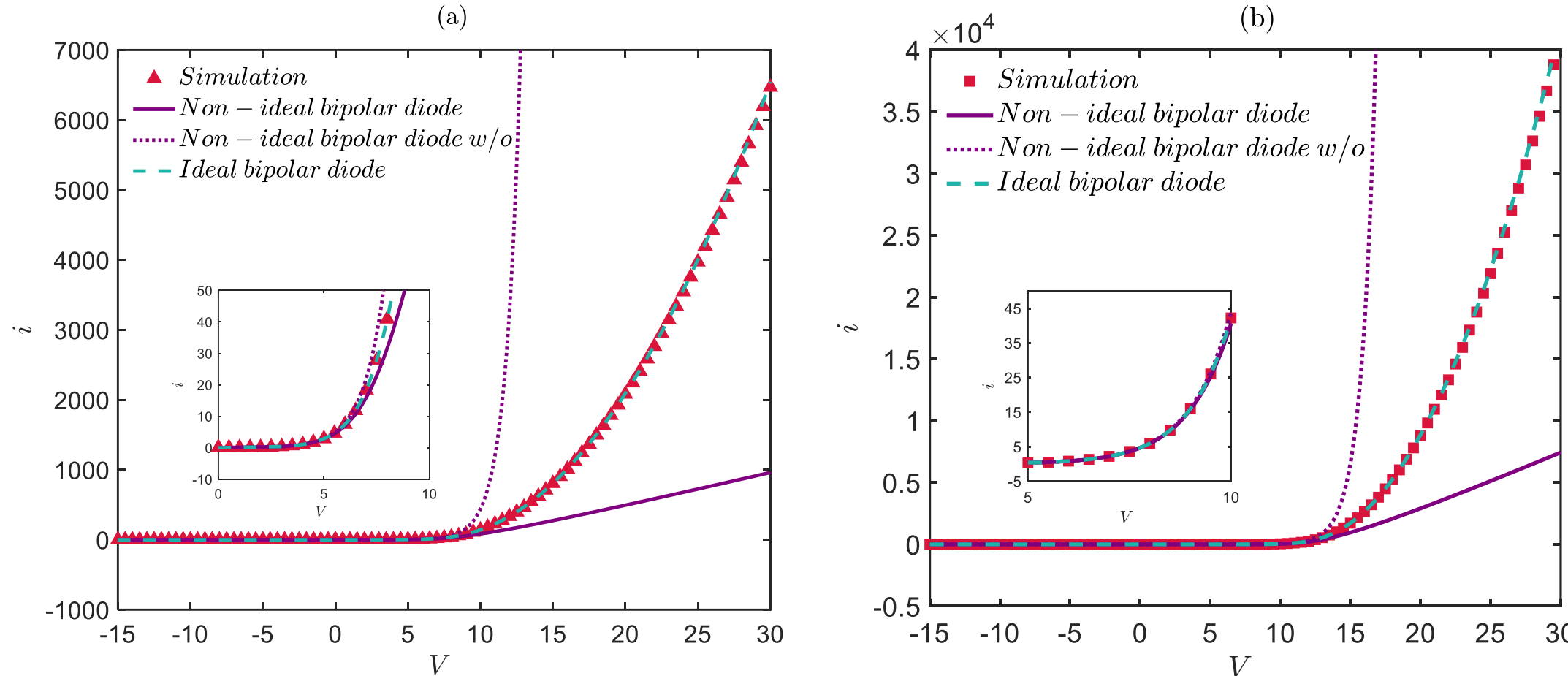

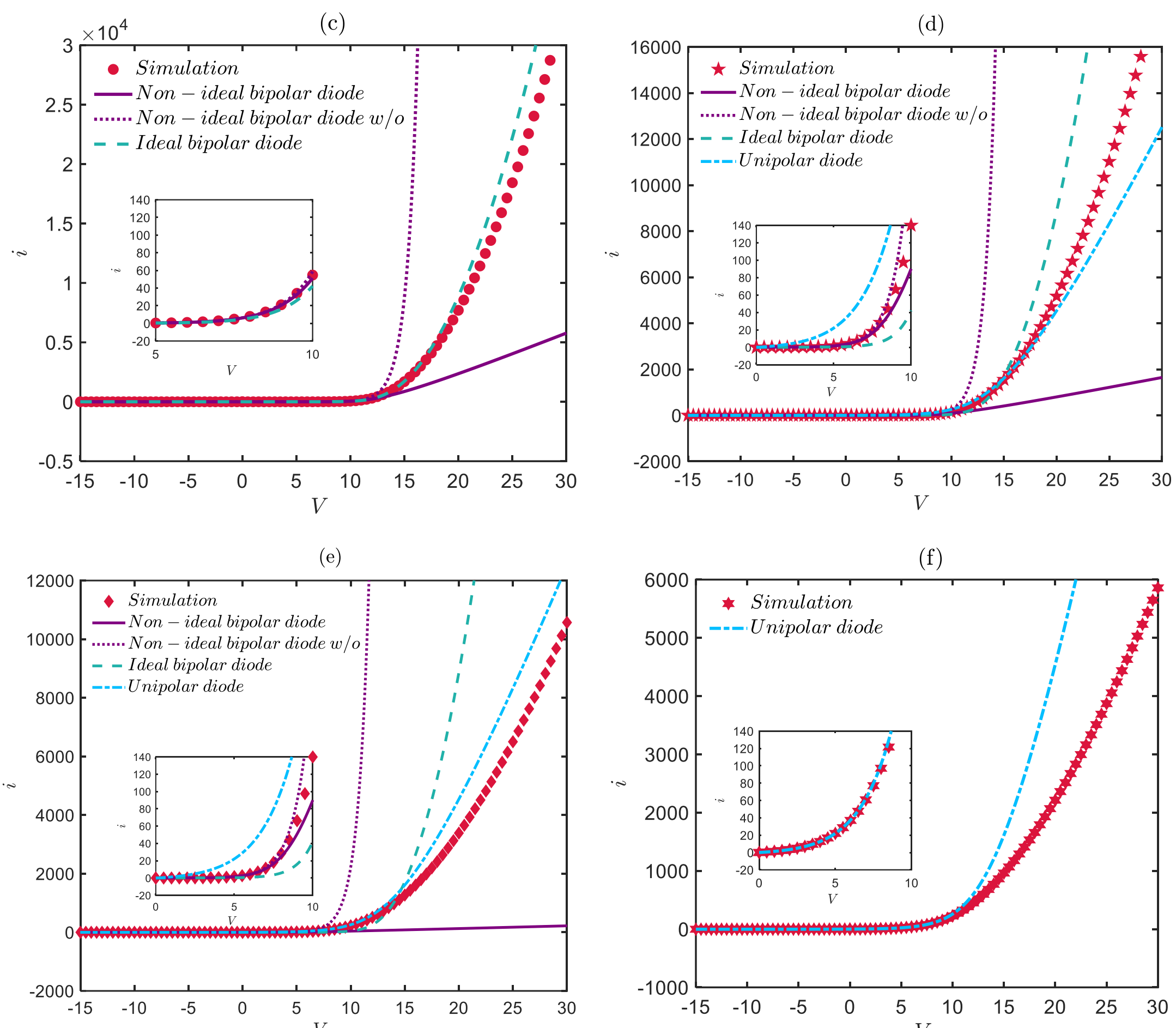

**Figure 4**. Current-density-voltage, $i-V$, curves corresponding to the scenarios given in the $N_1-N_2$ phase space diagram (Figure 3), where the simulation markers correspond to the values given in Table 2. The unipolar corresponds to Eq. (9), the ideal bipolar model corresponds to Eq. (10), and the non-ideal bipolar diode corresponds to Eq. (12), which is plotted with and without (w/o) the linear Ohmic term – see main text for discussion.

Figure 4(d) presents the results for $|N_2| = 100 \gg 1$. Here, we see that the non-ideal model has good correspondence at low voltages, while the correspondence is lost at high voltages. The loss of good correspondence now occurs at lower voltages (the explanation will be provided in Sec. 4.2). Also, in contrast to Figure 4(c), we see that the ideal model is no longer a good fit. However, this too can be rationalized, since $j \neq 0$. Importantly, we see that the expected result is that the curve is situated between the expected limits of unipolar and ideal diodes.

Figure 4(f) shows the result for $N_2 = 0$ $(\eta = 0)$, which corresponds to the case of a unipolar system. Here, we see that, like the non-ideal model, there is good correspondence at low voltages but poor correspondence at high voltages. It can be observed that the simulation lies underneath the predicted theoretical line. The origin of this behavior and the ultimate failure will be discussed below. Importantly, this failure will also explain why the simulation in Figure 4(e) for $N_2 = -10$ fall below the theoretical prediction.

**Preliminary summary**. We have shown that the simulated $i - V$ 's do not universally match a single model. This can be expected as the simulations depend on $N_1$ and $N_2$ in a manner that doesn't always correspond to the various models' dependence on $N_1$ and $N_2$. Moreover, it appears that there is also some sort of dependence on the applied voltage (or current) that is not captured by existing models. This will now be clarified.

### 4.2 Origin of high-voltage response failure

To understand the failure of the various models observed in Figure 4, one must return to the governing equations [Eq. (2)–(4)]. It can be observed that the concentrations and the potential depend on all system parameters (i.e., $N_1$, $N_2$, $L_1$, $L_2$), naturally leading to the $i - V$ 's dependency on them. In the following, we will delineate how the local concentration distribution and local electric potential distribution vary with the system parameters and applied voltage. We will show that when these distributions diverge from the assumed behavior, the theoretical models fail. We will present numerical simulations of the $c_\pm$ and $\phi$ for various values of $N_2$ and $V$ while keeping $N_1 = 1000$ and $L_1 = L_2 = 1$. Our discussion starts with $N_2 = -1000$ [Figure 5(a–c)], then we will jump to $N_2 = 0$ [Figure 5(g–i)], and complete the discussion with $N_2 = -100$ [Figure 5(d–f)].

#### 4.2.1 Ideal bipolar diode, $N_2 = -1000$

First, let us point out the observed symmetry of the concentration profiles [Figure 5(a)], where $c_+$ and $c_-$ in region 1 have perfect mirror symmetry with $c_-$ and $c_+$ in region 2, respectively. While the concentration profiles vary with the voltage, the mirror symmetry is voltage independent. This is an expected outcome of ideal bipolar systems, where $N_1 = |N_2|$. One can note that the difference between $c_+$ and $c_-$ in region 1, is identically equal (in the absolute sense) to the difference between $c_+$ and $c_-$ in region 2, indicating electroneutrality.

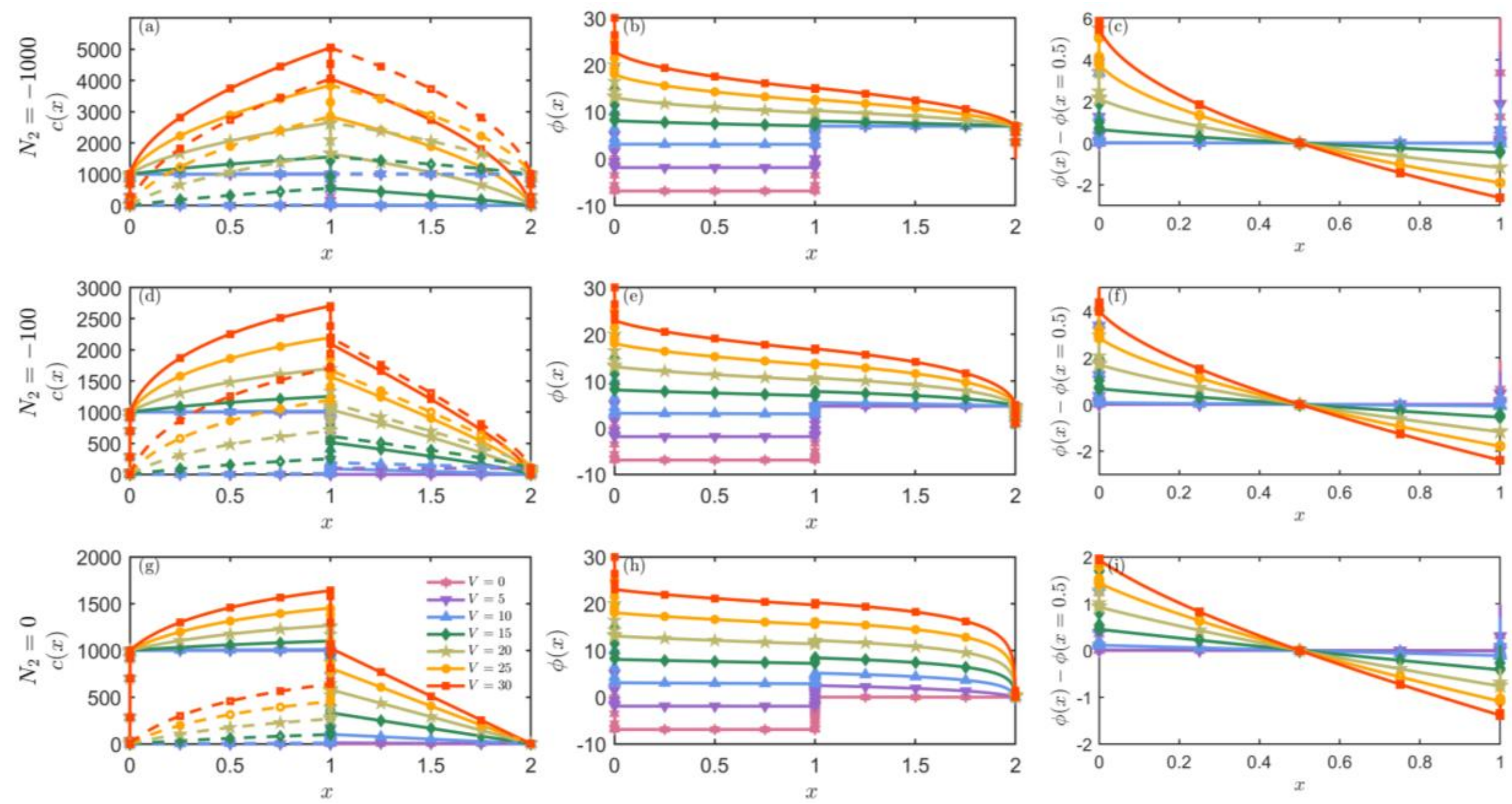

**Figure 5**. Numerically calculated spatial profiles of concentration and electric potential distribution for varying applied voltages. (a), (d), (g) Concentration profiles of positive (solid lines) and negative (dashed lines) ions as a function of the spatial coordinate. (b), (e), (h) Corresponding electric potential distributions $\phi(x)$ across the same spatial domain and voltage range. (c), (f), (i) Zoomed-in and shifted electric potential profiles, plotted as $\phi(x)-\phi(x=0.5)$ in region 1.

For the large voltages, it is rather clear that the profiles are not linear. In fact, they have a square-root dependency[40]. At low voltages, the profiles appear to be either constant or linear-like. However, this is just an approximate behavior in the same sense that the $P_k / Q_{\pm,k}$'s in Eq.(11) can be approximated to be constants or linear for very small and small currents, respectively. Similarly, the potential also has a square-root-like distribution[40] that can be observed at high voltages [Figure 5(b–c)]. At lower voltages, the potential appears to be constant or linear-like; however, here too this is to be expected (the exact dependency can be found in Refs.[40,50]). Another expected outcome[44] is that the electric potential is antisymmetric around the center of the channel, located at $x=1$ [Figure 5(b)].

#### 4.2.2 Unipolar diode, $N_2 = 0$

For low voltages ($V = 0,5,10$), we find the expected results that in region 1, $c_+ \approx N_1 = 1000$ and $c_- \approx 1/N_1 = 0.001$ [Figure 5(g)]. In region 2, $c_+ = c_- = O(1)$. Also, the potential in region 1 is approximately given by the Donnan potential, $-\ln(N_1)$ [Figure 5(h)], where the electric field in region 1 is very small [Figure 5(i)].

For intermediate voltages ($V = 15$), the positive concentration in region 1 ceases to be constant $c_+ \approx N_1$ and starts to increase [Figure 5(g)]. To satisfy electroneutrality, $c_-$ in region 1 must increase. The increase in both leads to an increase in the concentrations of both species in region 2 (the reason for this is explained in the next sub-section, Sec.4.3). In region 2, the concentrations are equal ($c_+ = c_-$) but are no longer of order unity. Now, the electric potential in region 1 is linear [Figure 5(h–i)].

For high voltages ($V \geq 20$), we observe that $c_+$ in region 1 is no longer constant or linear. To satisfy electroneutrality, the drastic increase in $c_+$ necessitates a drastic change of $c_-$, which leads to drastic change in $c_\pm$ in region 2 [Figure 5(g)]. The change in the behavior of $c_+$ and $c_-$ in region 1 also leads to a drastic change in the potential in this region [Figure 5(h)]. Instead of having an almost constant potential or even a linear drop, the potential drop is nonlinear [Figure 5(i)]. This is the origin of the failure observed in Figure 4(e–f), which underestimated the resistance of the system.

#### 4.2.3 Non-ideal bipolar diode, $N_2 = -100$

This scenario behaves like a hybrid of the other two scenarios. Let us first consider the concentration distributions shown in Figure 5(d). Similar to the two previous figures, both regions satisfy electroneutrality, whereby the difference between the positive and negative ions is always a constant ($N_1$ in region 1, $N_2$ in region 2). In region 1, the curves vary from a constant-like profile to a linear profile and then to a nonlinear spatially dependent profile (whose exact details are not known but likely resemble, to some degree, a square root profile). Yet, it can be observed that in region 2, for low voltages, the concentration profiles are almost linear [as in Figure 5(g)], but for high voltages, the profiles are square-root-like, as in Figure 5(a).

In contrast to the symmetric behavior of the electric potential observed in Figure 5(b) when $N_1 = -N_2$, when $N_1 \neq -N_2$, the response is not symmetric [Figure 5(e)]. Such an asymmetric response is more reminiscent of the unipolar diode shown in Figure 5(h). In Figure 5(f), we can observe the transition from an almost constant-like distribution to a linear drop (predicted by Ref.[52]), and then a nonlinear drop, as the voltages are increased.

**Preliminary summary**: While Sec.4.1 focused on showing that the theoretical current-density-voltage responses deviate from numerical simulations. Here, we focus on rationalizing why the oversimplification of the behavior of the concentration and electric potential distributions of the theoretical models leads to their eventual failure. In the next sub-section, we will explain why these changes occur and how the changes in region 1 affect region 2 (and vice versa).

### 4.3 Donnan potential

To understand the failure of all the models and the changes in the concentration and electric potential distributions, the effects of the Donnan potential drop must be discussed. To understand the effects of the Donnan potential, first consider the sharp changes occurring at the interface located at $x=1$. It almost appears as though there is concentration and potential discontinuity. However, in reality, there is no discontinuity – these changes occur over an extremely small layer – this is the (normalized) Debye length, $\varepsilon$ (whereby $\varepsilon \ll 1$). The Debye length is a microscopic length scale in which very sharp but continuous changes take place. Thus, on the macroscopic level, it appears that there is a jump. We will briefly describe how to model the physics of the jump in the distribution and how this jump is associated with the behaviors observed in Figure 5.

In numerical simulations, all the concentrations and the potential distributions are continuous, as are the derivatives from which one calculates the 1D fluxes ( $j_+$ and $j_-$ ) [and the currents ($i$ and $j$)]. If the fluxes ( $j_+$ and $j_-$ ) are derivatives of the electrochemical potential[63,64,54] given by $\mu_\pm = \ln(c_\pm) \pm \phi$, then $\mu_\pm$ must also be continuous. Thus, the sum of $\mu = \mu_+ + \mu_- = \ln(c_+ c_-)$ is continuous everywhere in the system, even at $x=1$.

In Figure 6, we plot $c_+$ and $c_-$ at $x=1^+$ and $x=1^-$, which are two points just to the right and left of $x=1$, respectively. Let us start with the ideal bipolar case, when $N_1 = -N_2$ shown in Figure 6(a). The mirror symmetry [from Figure 5(a)] ensures that $c_+(x=1^-) = c_-(x=1^+)$ and $c_-(x=1^-) = c_+(x=1^+)$ such that $c_+(x=1^-)c_-(x=1^-) = c_+(x=1^+)c_-(x=1^+)$ is indeed satisfied. Two things can be observed. At low voltages, the interfacial concentrations are almost constant, and at higher voltages, they increase with the voltage. Also, due to electroneutrality, the difference (in the absolute sense) in each region is $N_1$ (or $|N_2|$), respectively.

Figure 6(c) for the unipolar system exhibits similar but not identical behavior. From the point of view of conservation of electroneutrality, in region 2, where $N_2 \equiv 0$, we find the expected results that $c_+ = c_-$ for all voltages. In region 1, electroneutrality is also always conserved, but with a voltage dependency. At low voltages, we see the expected results (as asserted by Ref.[51]): $c_-(x=1^-)\approx 0$ and $c_+(x=1^-)\approx N_1$. At higher voltages, the interfacial concentrations become voltage dependent. Since $c_+(x=1^-)$ increases, conservation of electroneutrality requires that $c_-(x=1^-)$ as well. To satisfy $c_+(x=1^-)c_-(x=1^-)=c_+(x=1^+)c_-(x=1^+)$, the concentration in region 2, which is still maintained, increases (even to very large values of the order of $N_2$). Thus, the underlying assumption of the unipolar model of $c_+=N_1$ becomes void, yielding incorrect predictions for the $i-V$. This is the reason for the failure of the $i-V$ shown in Figure 4(e–f).

In Figure 6(b), we see a similar set of behaviors. Electroneutrality in regions 1 and 2 requires that the concentration differences are (in the absolute sense) $N_1$ and $N_2$, respectively. Here, too, we see that concentrations are voltage dependent. Importantly, to ensure that the multiplication of $c_+c_-$ are the same in both regions, then region 1 affects region 2 (and vice versa).

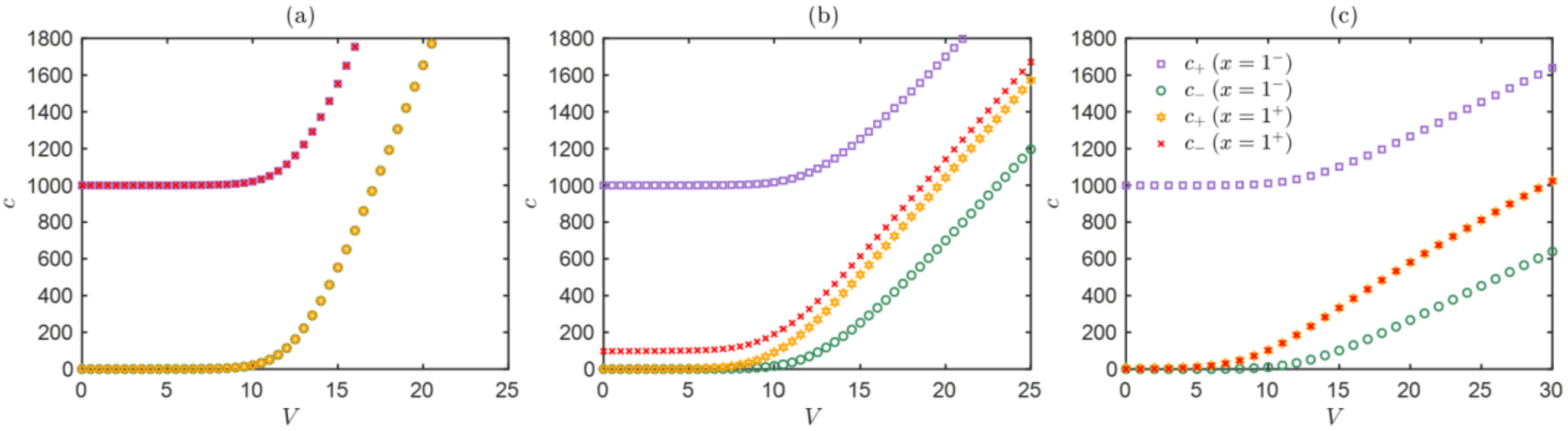

**Figure 6**. Interfacial concentrations (at $x=1$) vs. applied voltage $(c_\pm - V)$, with $N_1=1000$, and (a) $N_2=-1000$, (b) $N_2=-100$, and (c) $N_2=0$. The concentrations are evaluated at positions adjacent to the left and right of the interface at $x=1^\pm=1\pm\varepsilon$.

**Preliminary summary**: Here, we showed how the requirement that the multiplication of $c_+c_-$ is continuous, combined with the requirement that electroneutrality is conserved, leads to changes in the behavior of the interfacial concentrations. The changes in the interfacial concentrations drive the changes of the concentration and electric potential distributions in both regions.

### 4.4 System Selectivity

Thus far, we have considered the $i-V$ response for only six scenarios. However, numerical simulations were conducted for a much broader parameter set. For $N_1 = 1000$, we scanned both $N_2$ and $V$. The findings are summarized in Figure 7 and discussed now.

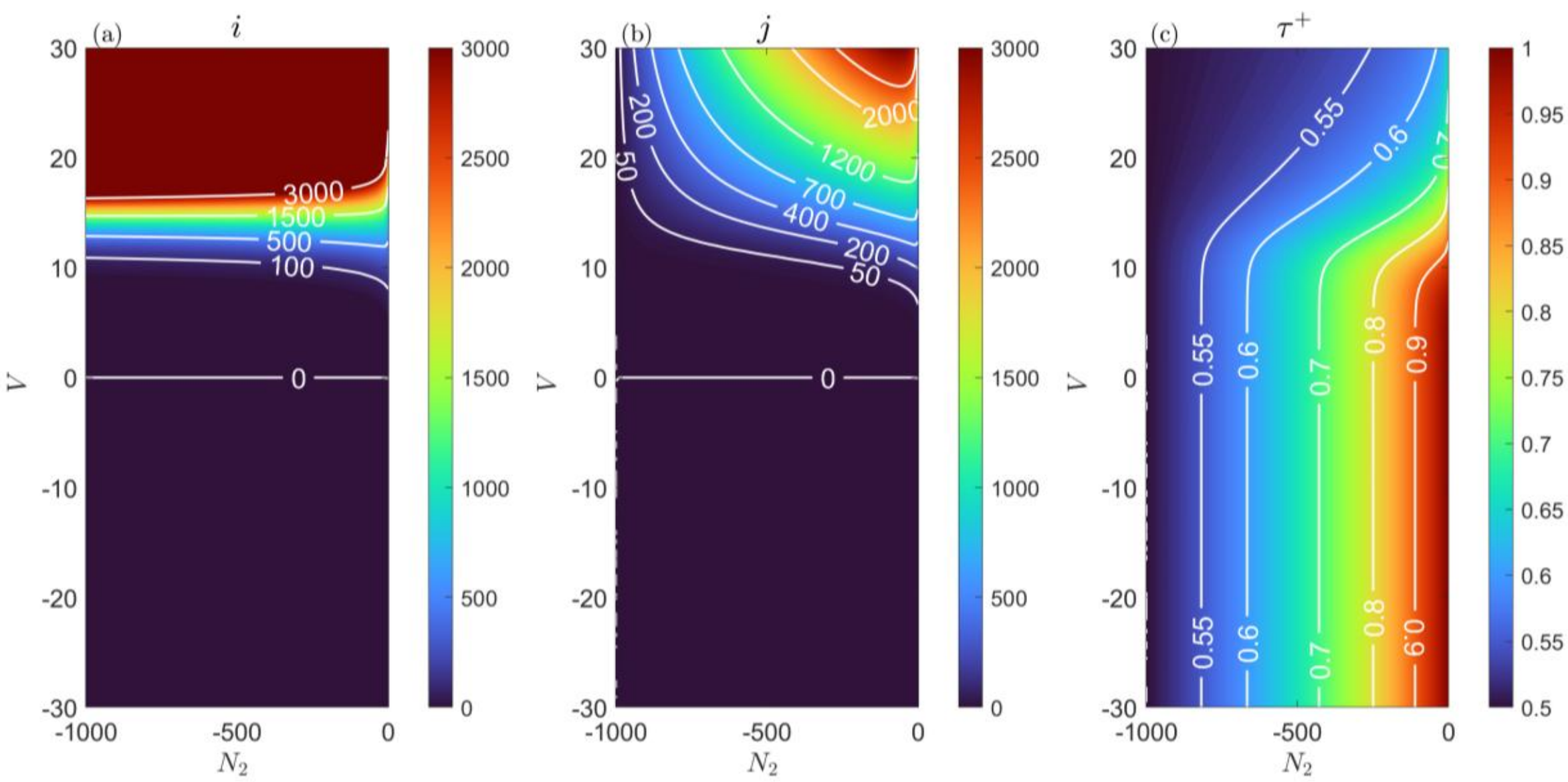

**Figure 7**. Numerically calculated (a) electric current density, $i$, (b) salt current density, $j$, and (c) positive ions transport number, $\tau^+$, versus the applied voltage $N_2$ and $V$ for fixed $N_1 = 1000$.

As has been emphasized throughout this work, the current $i$ increases with unrestricted growth for positive voltages but saturates to a small limiting current for negative voltages. The exact values, naturally, depend on the values of $N_2$ and $V$. Figure 7(a) presents this dependency, where it can be observed that there is a relatively stronger dependence on $V$ than on $N_2$. However, as we have discussed, this dual dependency is the reason that models eventually fail. In fact, we have shown that the theoretical models fail because they are unable to fully capture the changes in $j_+$ and $j_-$ (or $j$) due to changes in $N_2$ and $V$.

Figure 7(b) shows the behavior of $j$. It can be observed that $j$ is near zero or very small for much of the $N_2$ and $V$ phase space. However, the strong dependency on both $N_2$ and $V$ is even more prominent than it is for $i$. This strong dependency should no longer come as a surprise. We have established that the theoretical models fail because they make oversimplifying assumptions regarding the distributions of the concentrations and electric fields. These assumptions also oversimplify the behavior of the fluxes, such that the change is primarily in $j$ with smaller or even slight changes in $i$.

Two last comments are needed regarding the behavior of $i$ and $j$. First, Picallo *et al.*[52] noted their $i-V$ [Eq. (12)] failed at high voltages. Indeed, this is correct. In this work, we show that the failure for a given $N_1$ is not just voltage-dependent, but is also $N_2$ dependent. In fact, one can make a more general statement (for any scenario where $N_2 \neq 0$) – any theoretical model will fail for sufficiently large $V$, where the exact value will depend on $N_1$ and $N_2$.

Second, the change in the behavior of $j$, results in the change of the system's selectivity, which can be characterized by the transport number of either the positive species[41,56,44,59].

$$\tau^+ = \frac{j_+}{i} = \frac{1}{2}\left(1+\frac{j}{i}\right). \tag{13}$$

or the negative species $\tau^- = 1-\tau^+$. $\tau^+$ ranges from ½ to 1. When it is ½, the system is vanishingly selective, and when it is 1, the system is ideally selective.

In principle, for unipolar systems, the concept of selectivity is rather simple – it is the ability of the charged region to filter out ions of the opposite charge. This can be translated into a simple mathematical statement that if the co-ion concentration is small and/or the low co-ion flux is small, the system is highly selective ($\tau^+ \approx 1$). For unipolar systems, the selectivity of the system and the selectivity of the selective region are virtually indistinguishable. However, once bipolar systems are considered, the concept of selectivity becomes more complicated and the selectivity of the nanochannel and the system is distinctly different[44]. This should not come as a surprise. We have already shown that each region can be highly selective at equilibrium ($V=0$). Yet at sufficiently high voltages, the internal concentration polarization results in non-equilibrium distributions that reduce each region's selective capabilities. Thus, one needs to distinguish between the selective capability of each region and the total selectivity of the system[44].

Figure 7(c) shows that for bipolar systems, in general, even when $N_1 \gg 1$ and $|N_2| \gg 1$, such that each region is highly selective, the system is not highly selective. The system starts exhibiting higher selectivity as $N_2$ becomes smaller. In other words, as $N_2$ becomes smaller, and the system behaves more like a unipolar system, we see that the system's selectivity behaves like region 1's selectivity. However, even here, the unipolar model fails at high $V$.

This is because of the internal concentration polarization, which results in the transport of negative ions such that $j_-$ and $j$ are no longer zero.

**Preliminary summary**: The concept of selectivity in bipolar systems, comprised of two oppositely charged regions, is not straightforward. Each region can be highly selective, even when the system is not[44].

## 5. Discussion, future work, and conclusions

In this work, we have numerically investigated the $i-V$ response of nanofluidic bipolar systems. We have shown that the derived $i-V$'s (Table 1) are highly sensitive to the system's parameters through $\eta$ [Eq. (8)], which depends on the geometry and the excess counterion concentration of each region. From a naïve perspective, one would expect that the $i-V$ would only depend on $\eta$ such that, for a set geometry $(L_1 = L_2)$, the $N_1 - N_2$ phase space [Figure 2(a)] would be sufficient to fully characterize the electrical response. However, as we have shown here, this is not the case. The reason for this is the oversimplifying assumptions that are embedded in each of the analytic models. For each model, the assumption and its failure are different. The ideal bipolar model specifically assumes that $\eta = 1$. However, this is not truly attainable for a realistic experimental setup. Thus, any deviation from $\eta = 1$, will result in diminishing correspondence between theory and experiments. Similarly, the non-ideal bipolar diode and the unipolar assume that the concentrations in each region are almost constant and ignore the internal concentration polarization occurring within the system. As a result, they over- or under-predict the current (depending on the model).

### 5.1 Future works

#### 5.1.1 Extended phase space

We have shown thus far that the simple $N_1 - N_2$ phase space diagram [Figure 2(a)], while helpful, it is insufficient in providing a comprehensive approach to subdividing the $i-V$ response of all the systems. The reason is that it is over-simplified (as it is based on oversimplifying assumptions). Thus, we believe it is essential to extend the phase space to 3D – this is the extended $N_1 - N_2 - V$ phase space shown in Figure 8. The non-marked white regions represent the regions that cannot be identified with currently existing analytical models. Thus, if previously, it appeared that the three models spanned a non-trivial area of the $N_1 - N_2$ phase space, it now appears that they span even a relatively smaller volume of the extended phase space. This suggests the need to derive a more universal $i-V$ model. Until that universal model is derived and closes the knowledge gap presented here, it is also essential to provide experimentalists with several suggestions on how best to interpret their measurements (this will be discussed in Sec. 5.2).

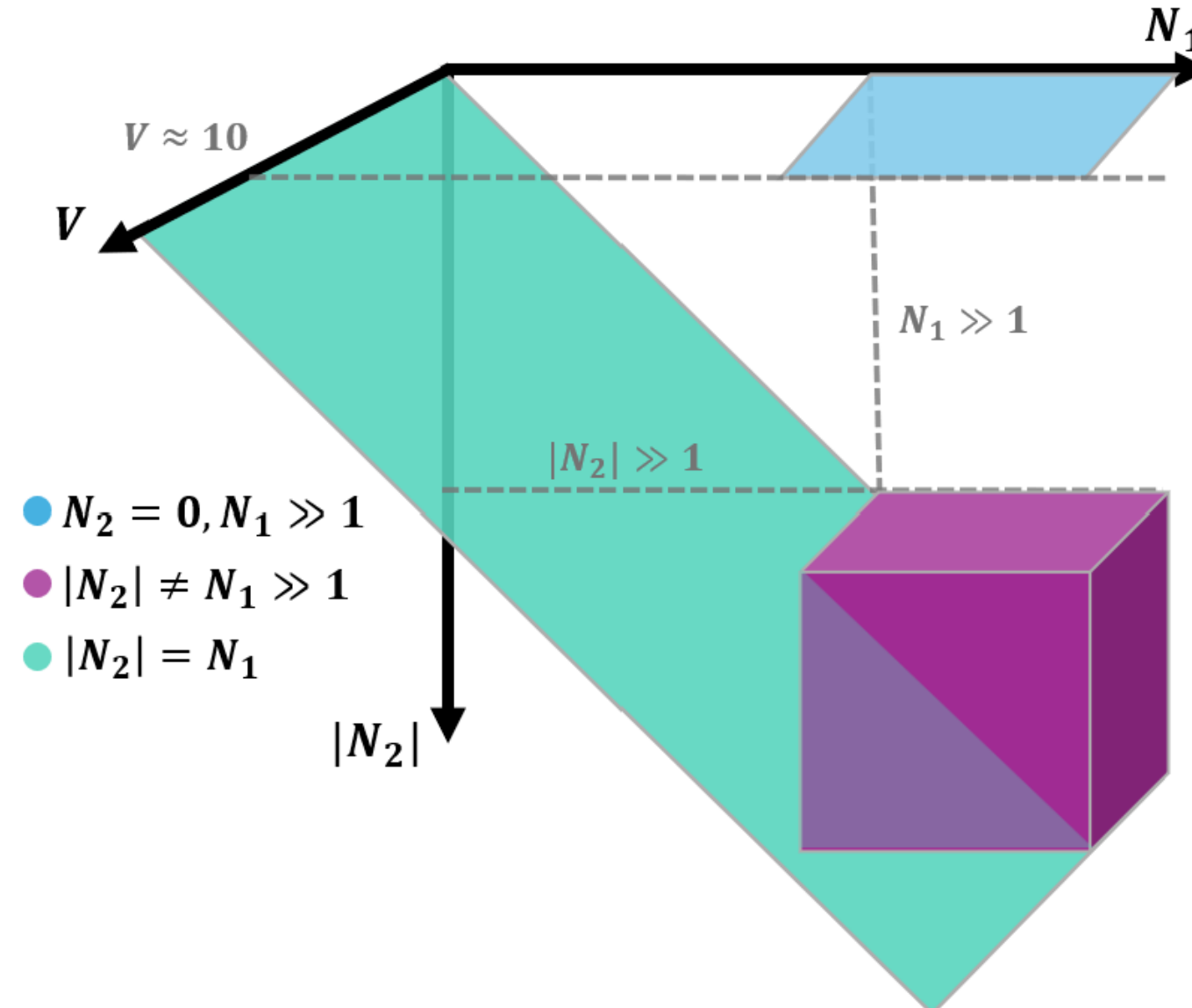

**Figure 8**. The extended $N_1 - N_2 - V$ phase space diagram denoting the parameter space where the three known current-density-voltage, $i-V$, responses for bipolar nanofluidic systems hold. Most of the phase space remains to be identified with one model or the other.

#### 5.1.2 Advection

In this work, we have assumed that advection is negligible. Since, thus far, we have presented only results from numerical simulations, in principle, this oversimplifying assumption can be removed and advection can be accounted for. However, the addition of advection carries with it non-trivial physical complications, as well as numerical complications. Our focus and discussion will be only on the physics.

It is trivial to add to Eqs. (2)–(4), the advective term $uc_{\pm}$ (in non-dimensional form, one needs to also multiply this by the Peclet number, which will be ignored for now – see Refs.[45,65] for more information). In principle, in a 1D system, the velocity (independent of whether it is pressure or electric field-driven) is a fitting parameter that cannot be calculated self-consistently[66,67]. For a pressure-driven flow, one can always use Darcy's law. However, this introduces additional complexities and fitting parameters.

To remove the dependency on any fitting parameter, one then needs to add the Stokes equations for the 2D or 3D vector $\mathbf{u}c_{\pm}$, as well as to add the electric body force. Not only is the mathematical complexity (and thus difficulties with numerical implementation) increased, but the resultant solution will then be dependent on the microscale geometry as well as the many different boundary conditions that can now be selected[65,68].

In all of the above arguments, we have assumed that the convection is forced into the nanoporous permselective region and the advection current is in the longitudinal direction. However, in principle, if the electrodes are not placed precisely on the two permselective materials, then an electroconvective instability[41] can form at one or both of the outer interfaces[45,61], yielding over-limiting currents (OLCs). OLCs due to advection occur for positive voltages (what is often termed "forward bias"). Note that these OLCs are distinctly different than OLCs that occur due to water-splitting discussed in the next subsection.

Regardless, as apparent from the above discussion, the issue of advection is not a trivial question and will require several works, some theoretical and some numerical, to address in a coherent manner. Importantly, this work, with its emphasis on advection-less transport, will be able to provide the required insights on how to analyze the additional contribution of advection.

#### 5.1.3 Four species

In this work, we considered an electrolyte comprised of two species. In reality, there are always more species. To the best of our knowledge, the general problem of an arbitrary number of species does not yield a tractable solution that allows a simple, straightforward analysis. However, one can consider a "reduced" problem, which is limited to three or four species. Perhaps the easiest scenario to consider is the following four species electrolyte comprised of two major charge carriers(like KCl considered in this work) and two minor charge carriers, which are typically the products of water dissociation ($OH^-$ and $H^+$)[69].

Recently, Ref.[34] considered the response of such a system. There, the authors had a KCl-like electrolyte (where the diffusion coefficients were equal) as the major charge carriers and $OH^-$ and $H^+$ (which also had equal diffusion coefficients) as the minor charge carriers. Ref.[34] then proceeds to analyze their system under the assumption of $j=0$, such that their system is equivalent to what we have termed the ideal bipolar scenario. Their numerical results show that the minor charge carriers do not contribute to the $i-V$ for positive voltages. However, at sufficiently large (in the absolute sense) negative voltages (what is often termed "reversed bias"), OLCs appear due to the additional contribution of the minor charge carriers.

In principle, one can now consider systems which are "extensions" to the formulation presented in Ref.[34], whereby several "internal symmetries" can be removed. In particular, one does not need to assume that the system is ideal (whereby $\eta=1$). However, one can also remove the assumption of asymmetric diffusion of either or both of the major/minor charge carriers. The results presented in this work will serve as the baseline response of the more general four-species response.

### 5.2 Experimental interpretation

To truly characterize a system, $L_1$, $L_2$, $N_1$ and $N_2$ need to be known. In principle, measuring $L_1$ and $L_2$ is a straightforward process that can, should, and must be measured during the pre-measurement stage (i.e., fabrication process). In what follows, we will assume that $L_1$ and $L_2$ are known.

We can consider the situation where $N_1$ and $N_2$ are known. In this scenario, one does not need any curve fitting. One can insert the known values into all three models and find which is the most appropriate. Naturally, one will not achieve perfect correspondence, but at least it will provide the "best" or most appropriate framework.

It is important to state that independent measurements of $N_1$ and $N_2$ is actually a straightforward, but rather time-consuming process. The simplest way possible to measure these quantities is to fabricate two systems that are comprised of only one charged region (once negative with $N_1$, and once positive with $N_2$). Then, one needs to undertake the established method of measuring the conductance versus the bulk concentration to extract $N_1$ or $N_2$ [41,65,70–76]. Measuring both $N_1$ and $N_2$ (independently) is the preferred route. But even one of them would provide critical information.

Suppose that the above suggested characterization of the surface charge density (or excess counterion concentrations) has yet to be conducted, such that $N_1$ and $N_2$ are unknowns. Then the question arises, which is the most appropriate model to use? As discussed, the models will eventually fail at high positive voltages, and therefore, when curve fitting the data, one is limited to relatively low voltages. We suggest limiting the voltages to approximately $V=5$, which is $\tilde{V}=0.125\,[V]$ in dimensional units. Then, the $i-V$ should be experimentally measured for 3–4 different concentrations. The lowest concentration should ensure that $N_1 \gg 1$ while the highest concentration should ensure that $N_1 \ll 1$. Then, the information regarding the conductance and limiting currents provided in Table 1 can be utilized.

One should remember two important facts. The ideal bipolar model holds for all values of $N_1=-N_2$ which can be small or large, corresponding to high or low concentrations, respectively. Such that if one satisfies $N_1=-N_2$, then a limiting current will always appear.

This is in contrast to the unipolar model and the non-ideal bipolar that hold only for low concentrations when $N_1 \gg 1$. For intermediate and high concentrations, these models can no longer predict a limiting current. Thus, if one finds that for all the 3–4 considered concentrations (especially when $N_1 \ll 1$ and $|N_2| \ll 1$), the $i-V$ still exhibits a limiting current, then the only model that is still appropriate is the ideal bipolar model. From the dimensional limiting current, $\tilde{i}_{\lim} = 2\tilde{F}\tilde{D}/(\tilde{L}_2\tilde{N}_2)$, which is concentration-independent, one can extract $\tilde{N}_2$ (which is also $-\tilde{N}_1$).

If the limiting current vanishes at high concentrations ($N_1 \ll 1$), then we are left to consider the unipolar model and the non-ideal bipolar. Their limiting currents are given by $\tilde{i}_{\lim} = -2\tilde{F}\tilde{D}\tilde{c}_0/\tilde{L}_2$ and $\tilde{i}_{\lim} = \tilde{F}\tilde{D}(\tilde{N}_1\tilde{L}_1 - \tilde{N}_2\tilde{L}_2)/(\tilde{N}_2\tilde{L}_2\tilde{N}_1\tilde{L}_1)$ respectively. It can be observed that the former is concentration-dependent while the latter is concentration-independent. Then, if two $i-V$'s are measured at two different bulk concentrations that satisfy $N_1 \gg 1$, one can test whether the ratio of the two measured limiting currents depends on the bulk concentration ratio.

Our finding that the three classical models break down at high voltages may or may not be surprising. Yet to date, to the best of our knowledge, the breakdown has not been reported and thoroughly discussed. The breakdown underscores the need to derive a more comprehensive framework to capture a broader range of system behaviors. However, the findings of this work also provide a robust framework for analyzing the electrical response of nanofluidic systems. This framework can be used to guide the design of more efficient nanofluidic devices as well as to enhance the analysis of experimental measurements.

**Associated Content**

**Methods.** *Comsol numerical simulations*. The PNP equations [Eqs.(2)–(4)], subject to the boundary conditions [Eq. (6)] were solved using the finite elements program COMSOL™, for the one-dimensional geometry described in Figure 1, using the Transport of Diluted Species and Electrostatic modules. A detailed discussion on our numerical methodology can be found in our recent works[44,45,61].

In addition to the values given in Table 2, we used $\varepsilon = 3\cdot 10^{-4}$. To resolve the Donnan potential jumps at the three interfaces, we used a highly refined mesh of approximately $10^4$ elements, spanning a distance of $2\varepsilon$ from the interface. Elsewhere, a relatively coarse mesh was used. To reduce the run time and ensure convergence, the initial conditions/guesses used were $c_{+,1} = N_1$, $c_{-,1} = 1/N_1$, $\phi = -\ln(N_1)$ and $c_{-,2} = |N_2|$, $c_{+,2} = 1/|N_2|$, $\phi = \ln(|N_2|)$ in regions 1 and 2, respectively. For the $i-V$'s, we scanned $V$ from $-30$ to $30$ with steps of $dV=0.5$ and $N_2$ from $-1000$ to $0$ with steps of $dN=10$ for two scenarios, where $N_1=100$ and $N_1=1000$.

**Data availability statement.** The data that support the findings of this article are not publicly available. The data are available from the authors upon reasonable request.

**Competing interests.** The authors declare that they have no conflicts of interest.

**Funding.** This work was supported by Israel Science Foundation grants 204/25.

**Acknowledgements.** We acknowledge the support of the Ilse Katz Institute for Nanoscale Science & Technology and the Pearlstone Center for Aeronautical Engineering Studies. Additionally, A.B-K.S. was supported by the Ministry of Energy and Infrastructure's scholarship program for graduate students in the fields of energy, earth, and marine sciences.

**Author contributions.** Ayelet Ben-Kish Sharvit: conceptualization, investigation, validation, writing, and editing. Yoav Green: conceptualization, investigation, resources, supervision, writing, and editing.